\documentstyle[11pt,newpasp,twoside,epsf]{article}
\def\emphasize#1{{\sl#1\/}}

\begin{document}
\title{The evolution of radio galaxy emission line regions: Breaking
the P-z degeneracy}
\author{K. J. Inskip, M. S. Longair}
\affil{Astrophysics Group, Cavendish Laboratory, Cambridge, CB3 0HE}
\author{P. N. Best}
\affil{Institute for Astronomy, Edinburgh, EH9 3HJ}

\begin{abstract}
The ionization and kinematic properties of the emission line regions
of three subsamples of 6C and 3CR radio galaxies have been compared.
The degeneracy between redshift and radio power is broken, and the
relative importance of radio power (P), radio size (D) and redshift
(z) for the emission line region properties is determined.

\end{abstract}

\section{Introduction}
The emission line regions (ELRs) of high redshift radio galaxies are
generally more extensive than those of lower redshift
sources.  Spectroscopic studies of 3CR sources at $z \sim 1$
(Best, R\"{o}ttgering \& Longair 2000a, 2000b) reveal that the size of
the radio source plays an 
important role in determining the dominant ionization mechanism.  The
emission lines of small sources are best explained by the predictions
of shock ionization associated with the passage of the radio cocoon.
Larger sources usually appear photoionized by the central AGN, and
typically exist in a higher ionization state than their smaller
counterparts. 

Studies of 3CR radio galaxies reveal considerable changes with
redshift, with the lower redshift/radio power sources (degenerate in
the flux limited 3CR sample) generally having much less extreme
kinematics than sources at $z \sim 1$ (Baum \& McCarthy 2000; Best et
al 2000b).  This degeneracy can be broken by means of a comparison
with other radio galaxy surveys, e.g. the 6C sample.

\section{Breaking the redshift--radio power degeneracy}
We have studied a complete subsample of 8 6C galaxies within the
redshift range $0.85 < z < 1.3$ (Inskip et al, {\it in prep}).   
These are a factor of $\sim6$ lower in radio power than 3CR sources at
the same redshift.  The ionization and kinematic properties of the
ELRs of these galaxies have been compared with those
of the $z \sim 1$ 3CR subsample of Best et al (2000a), and also with
lower redshift 3CR sources matched in radio power to the 6C $z \sim 1$
subsample.  Low redshift ELR kinematic data were obtained
from Baum \& McCarthy (2000); ionization data were obtained from
Tadhunter et al (1998).  This combination of samples allows us to
break the P--z degeneracy effectively.    

\begin{figure}
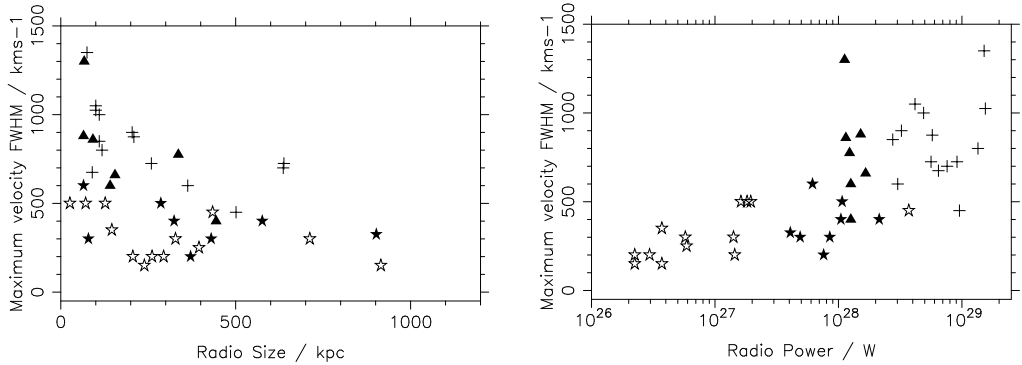

\plotfiddle{Fig_Drad.eps}{80.pt}{-90.}{28.}{28.}{-200.}{125.}
\plotfiddle{Fig_Lrad.eps}{0.pt}{-90.}{28.}{28.}{0.}{150.}
\caption{These plots display the variation in maximum velocity FWHM
with (a - left) D and (b - right) P for three radio galaxy
subsamples. 3CR sources at $z \sim 1$ are denoted by crosses, and 6C
sources at $z \sim 1$ by filled triangles. The stars denote the low
redshift 3CR data. (Filled stars represent the low redshift 3CR
sources with comparable radio powers to the 6C subsample.)}
\end{figure}

Fig. 1 displays kinematic data for three subsamples, plotted against
D and P.  Principal component analysis and partial rank correlation
tests on these data sets have been used to determine the relative
importance of P, D and z on the changing ionization and kinematic
properties of radio galaxy ELRs.  For larger radio galaxies, the gas
kinematics become less extreme, and the ionization state increases, as
has been found in previous studies (e.g. Best et al 2000b).  In
addition to this, we also find that ionization state is strongly
dependent on P but \emphasize{not} z, with the more powerful radio
sources existing in higher ionization states than less powerful
sources independent of redshift.  On the other hand, the FWHM of the
ELRs is correlated most 
strongly with redshift, as well as with radio power. The most extreme
kinematics are observed in powerful radio sources at high redshift. 

From this analysis we conclude that: \\
1. AGN properties vary little with redshift, and the power/age of a
radio source is the determining factor in the ionization mechanism of
the emission line gas.  \\
2. The gas kinematics points to significant evolution of the host
galaxy and ELR properties with redshift, in addition
to their dependency on the properties of the radio source.

\end{document}